# COUPLING HIDDEN MARKOV MODELS FOR THE DISCOVERY OF CIS-REGULATORY MODULES IN MULTIPLE SPECIES

By Qing Zhou and Wing Hung Wong[1]

*UCLA and Stanford University*

*Cis*-regulatory modules (CRMs) composed of multiple transcription factor binding sites (TFBSs) control gene expression in eukaryotic genomes. Comparative genomic studies have shown that these regulatory elements are more conserved across species due to evolutionary constraints. We propose a statistical method to combine module structure and cross-species orthology in *de novo* motif discovery. We use a hidden Markov model (HMM) to capture the module structure in each species and couple these HMMs through multiple-species alignment. Evolutionary models are incorporated to consider correlated structures among aligned sequence positions across different species. Based on our model, we develop a Markov chain Monte Carlo approach, MultiModule, to discover CRMs and their component motifs simultaneously in groups of orthologous sequences from multiple species. Our method is tested on both simulated and biological data sets in mammals and Drosophila, where significant improvement over other motif and module discovery methods is observed.

**1. Introduction.** Gene transcription is regulated by interactions between transcription factors and their binding sites on DNA. The analysis of genomic sequences for short sequence elements (*cis*-regulatory elements) that mediate such interactions is an important problem in computational biology. In this paper we develop a method for predicting *cis*-regulatory elements based on the statistical modeling of combinatorial control by multiple transcription factors, and of cross-species conservation of the regulatory roles of these factors. The remaining part of this Introduction provides a review of relevant literature and background of our approach. Section 2 presents our

Received October 2006; revised February 2007.
[1]Supported by NIH Grants GM067250 and HG003903.
  Supplementary material available at http://imstat.org/aoas/supplements

statistical model. Section 3 develops the computational algorithms for inference from this model. Sections 4 and 5 present evidence for the effectiveness of the proposed method through both simulation studies and applications to genomic data. Section 6 contains concluding remarks and discussions of future work. The last section is an Appendix on mathematical details.

A transcription factor (TF) recognizes and binds to many different sites in the genome. The sites for a TF are usually not identical but share similarity that can be summarized by a statistical model called a motif. A motif is often parameterized by a position-specific weight matrix (PWM) that summarizes the relative frequencies for the four types of nucleotides at each position of the site. Computational methods for motif discovery and TFBS prediction were initiated by Stormo and Hartzell (1989) and further developed in Lawrence and Reilly (1990), Lawrence et al. (1993) and Liu, Neuwald and Lawrence (1995). An alternative approach based on word enumeration instead of PWM was proposed by Bussemaker, Li and Siggia (2000), Hampson, Kibler and Baldi (2002), Liu, Brutlag and Liu (2002) and Sinha and Tompa (2002).

A number of approaches have been developed recently to increase the accuracy of TFBS prediction based on the PWM formulation. The first approach utilizes the concept of *cis*-regulatory modules [Yuh, Bolouri and Davidson (1998), Loots et al. (2000)]. A *cis*-regulatory module is a short sequence element (typically 100–500 bp in length) containing a cluster of TFBSs for one or more TFs. Different TFs binding to the same *cis*-regulatory module cooperate in their control of gene transcription through mechanisms such as synergistic binding or sequential recruitment. Computational methods that search for CRMs given the PWMs of several interacting TFs were proposed by Frith, Hansen and Weng (2001), Frith et al. (2002) and Berman et al. (2002). Such searches were combined with the alignment of orthologous sequences from several related species [Sinha, van Nimwegen and Siggia (2003)]. When the PWMs are unknown, a *de novo* module discovery method has been developed using a hierarchical mixture model for the CRM [Zhou and Wong (2004)]. This method, called CisModule, has been applied successfully to predict the CRMs that control Ciona muscle development [Johnson et al. (2005)]. In addition, a model that considers transition probabilities and neighboring distances between TFBSs has been proposed in Thompson et al. (2004), and was further developed to identify CRMs given a collection of potential PWMs [Gupta and Liu (2005)].

Several recent methods employ a second approach, that is, multiple genome comparison, to enhance the power of *cis*-regulatory analysis. PhyloCon [Wang and Stormo (2003)] builds multiple alignments among orthologs and extends these alignments to identify motif profiles. CompareProspector [Liu et al. (2004)] biases motif search to more conserved regions based on conservation scores. OrthoMEME [Prakash et al. (2004)] identifies pairs of orthologous

TFBSs in two species. With a given alignment of orthologs and a phylogenetic tree, EMnEM [Moses, Chiang and Eisen (2004)], PhyME [Sinha, Blanchette and Tompa (2004)] and PhyloGibbs [Siddharthan, Siggia and Nimwegen (2005)] detect motifs based on more comprehensive evolutionary models for TFBSs. Finally, when evolutionary distances among the genomes are too large for the orthologous sequences to be reliably aligned, Li and Wong (2005) proposed an ortholog sampler that finds motifs in multiple species independent of ortholog alignments.

Modeling CRMs enhances the performance of *de novo* motif discovery because it allows the use of information encoded by the spatial correlation among TFBSs in the same module. Likewise, the use of multiple genomes enhances motif prediction because it allows the use of information from the evolutionary conservation of TFBSs in related species. Although both types of information have been exploited separately, they have not been utilized simultaneously for *de novo* prediction for *cis*-regulatory elements. We believe that an approach that utilizes both pieces of knowledge can further improve the power for *de novo* prediction. In this paper we use a hidden Markov model (HMM) to capture the co-localization tendency of multiple TFBSs within each species, and then couple the hidden states (which indicate the locations of modules and TFBSs within the modules) of these HMMs through multiple-species alignment. We develop evolutionary models separately for background nucleotides and for motif binding sites in order to capture the different degrees of conservation among the background and among the binding sites. We develop a Markov chain Monte Carlo algorithm for sampling CRMs and their component motifs simultaneously from their joint posterior distribution. We test the method on both simulated and well-annotated biological data sets, and demonstrate that it provides significant improvement over other *de novo* motif and module discovery methods. Compared to alignment-based motif discovery methods such as PhyME [Sinha, Blanchette and Tompa (2004)] and PhyloGibbs [Siddharthan, Siggia and van Nimwegen (2005)], our approach has two unique features: (1) We consider module information through a hidden Markov model; (2) The multiple alignments of orthologous sequences are dynamically updated, so that the uncertainty in the alignments is taken into account. The advantages of these features are illustrated by the examples in this article.

**2. The coupled hidden Markov model.** Our input data consist of upstream or regulatory sequences of $n$ (co-regulated) genes from $N$ species, that is, a total of $n \times N$ sequences. Assuming these genes are regulated by CRMs composed of binding sites of $K$ TFs, one wants to find these TFBSs and their motifs (PWMs). We only consider $N$ closely related species in the sense that their orthologous TFs share the same binding motif, which applies to groups of species within mammals, or within Drosophila, and so on.

Please note that our model, to be developed in this section, is applicable to the situation where different genes have distinct numbers of orthologs, that is, some orthologs are missing for some genes. For notational ease, missing orthologs are treated as zero-length sequences, so that we always assume $n$ sequences from each species.

2.1. *The HMM for module structure.* Let us first focus on the module structure in one sequence. We assume that the sequence is composed of two types of regions, modules and background. A module contains multiple TFBSs separated by background nucleotides, while background regions contain only background nucleotides. Accordingly, we assume that the sequence is generated from a hidden Markov model with two states, a module state ($M$) and a background state ($B$). In a module state, the HMM either emits a nucleotide from the background model (of nucleotide preference) $\theta_0$, or it emits a binding site of one of the $K$ motifs (PWMs) $\Theta_1, \Theta_2, \ldots, \Theta_K$. The probability for emission from $\theta_0$ and $\Theta_k$ ($k = 1, 2, \ldots, K$) is denoted by $q_0$ and $q_k$, respectively ($\sum_{k=0}^{K} q_k = 1$) [Figure 1(A)]. Note that a module state can be further decomposed to $K + 1$ states, corresponding to within-module background ($M_0$) and $K$ motif binding sites ($M_1$ to $M_K$), that is, $M = \{M_0, M_1, \ldots, M_K\}$. Assuming that the width of motif $k$ is $w_k$, a binding site of this motif, a piece of sequence of length $w_k$, is treated as one state of $M_k$ as a whole ($k = 1, 2, \ldots, K$). The transition probability from a background to a module state is $r$, that is, the chance of initiating a new module is $r$. The transition probability from a module state to a background state is $t$, that is, the expected length of a module is $1/t$. We denote the transition matrix by

$$(1) \qquad T = \begin{bmatrix} T(B, B) & T(B, M) \\ T(M, B) & T(M, M) \end{bmatrix} = \begin{bmatrix} 1 - r & r \\ t & 1 - t \end{bmatrix}.$$

This model can be viewed as a stochastic version of the hierarchical mixture model (HMx) defined in Zhou and Wong (2004).

2.2. *Coupling HMMs via multiple alignment.* The HMMs in different orthologs are coupled through multiple alignment, so that the hidden states of aligned bases in different species are collapsed into a common state [Figure 1(B)]. For instance, the nucleotides of state 4 in the three orthologs are aligned in Figure 1(B). Thus, these three states are collapsed into one state, which determines whether these aligned nucleotides are background or binding sites of a motif. (Note that these aligned nucleotides in different orthologs are not necessarily identical.) Here hidden states refer to the decomposed states, that is, $B$ and $M_0$ to $M_K$, which specify the locations of modules and motif sites. This coupled hidden Markov model (c-HMM hereafter) has a natural graphical model representation [lower panel of Figure 1(B)], in which each state is represented by a node in the graph and the arrows specify the dependence among them. The transition (conditional) probabilities for nodes with a single parental node are defined by the same $T$ in (1). We define the conditional probability for a node with multiple parents as follows: If node $Y$ has $m$ parents, each in state $Y_i$ $(i = 1, 2, \ldots, m)$, then we have

$$(2) \quad P(Y|Y_1, \ldots, Y_m) = \frac{1}{m} \sum_{i=1}^{m} P(Y|Y_i) = \frac{C_B}{m} T(B, Y) + \frac{C_M}{m} T(M, Y),$$

where $C_B$ and $C_M$ are the numbers of the parents in states $B$ and $M$, respectively ($m = C_B + C_M$). This equation shows that the transition probability to a node with multiple parents is defined as the weighted average from the parental nodes in background states and module states. We use the same emission model described in the previous section for unaligned states. For aligned (coupled) states, we assume star-topology evolutionary models with one common ancestor, although the method can be readily generalized to a tree topology. The c-HMM first emits (hidden) ancestral nucleotides by the emission model defined in Figure 1(A), given the coupled hidden states. Then, different models are used for the evolution from the ancestral to descendant nucleotides depending on whether they are background or TFBSs.

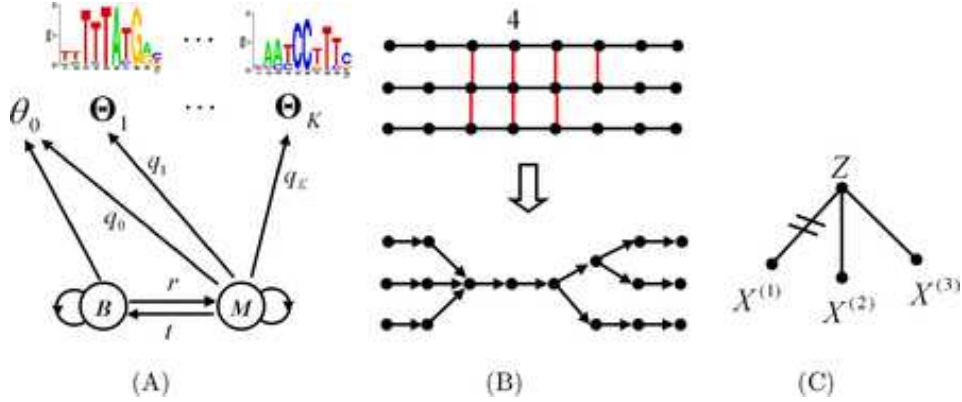

Fig. 1. *The coupled hidden Markov model (c-HMM).* (A) *The HMM for module structure in one sequence.* (B) *Multiple alignment of three orthologous sequences (upper panel) and its corresponding graphical model representation of the c-HMM (lower panel). The nodes represent the hidden states. The vertical bars in the upper panel indicate that the nucleotides emitted from these states are aligned and thus collapsed in the lower panel. Note that a node will emit $w_k$ nucleotides if the corresponding state is $M_k$ $(k = 1, \ldots, K)$.* (C) *The evolutionary model for motifs using one base of a motif as an illustration. The hidden ancestral base is $Z$, which evolves to three descendant bases $X^{(1)}$, $X^{(2)}$ and $X^{(3)}$. Here the evolutionary bond between $X^{(1)}$ and $Z$ is broken, implying that $X^{(1)}$ is independent of $Z$. The bond between $X^{(2)}$ and $Z$ and that between $X^{(3)}$ and $Z$ are connected, which means that $X^{(2)} = X^{(3)} = Z$.*

2.3. *The evolutionary model.* A neutral substitution matrix is used for the evolution of aligned background nucleotides, both within and outside of modules, with a transition rate of $\alpha$ and a transversion rate of $\beta$:

$$(3) \qquad \Phi = \begin{bmatrix} 1-\mu_b & \beta & \alpha & \beta \\ \beta & 1-\mu_b & \beta & \alpha \\ \alpha & \beta & 1-\mu_b & \beta \\ \beta & \alpha & \beta & 1-\mu_b \end{bmatrix},$$

where the rows and columns are ordered as $A$, $C$, $G$ and $T$, and $\mu_b = \alpha + 2\beta$ is defined as the background mutation rate. We assume an independent evolution for each position (column) of a motif under the nucleotide substitution model of Felsenstein (1981), which was also used in Sinha, van Nimwegen and Siggia (2003). Suppose the weight vector of a particular position in the motif is $\theta$. The ancestral nucleotide, denoted by $Z$, is assumed to follow a discrete distribution with the probability vector $\theta$ on $\{A, C, G, T\}$. If $X$ is a corresponding nucleotide in a descendant species, then either $X$ inherits $Z$ directly (with probability $\mu_f$) or it is generated independently from the same weight vector $\theta$ (with probability $1 - \mu_f$). The parameter $\mu_f$, which is identical for all the positions within a motif, reflects the mutation rate of the TFBSs. This model takes PWM into account in the binding site evolution, which agrees with the nonneutral constraint of TFBSs that they are recognized by the same protein (TF). It is obvious that, under this model, the marginal distribution of any motif column is identical in all the species. This evolutionary model introduces another hidden variable which indicates whether $X$ is identical to or independent of $Z$ for each base of an aligned TFBS. We call these indicators evolutionary bonds between ancestral and descendant bases [Figure 1(C)]. If $X = Z$, we say that the bond is connected; if $X$ is independent of $Z$, we say that the bond is broken.

### 3. Gibbs sampling and Bayesian inference.

3.1. *Basic framework.* Our full model involves the following parameters: the transition matrix $T$ defined in equation 1, the mixture emission probabilities $q_0, q_1, \ldots, q_K$, the motif widths $w_1, \ldots, w_K$, the PWMs $\Theta_1, \ldots, \Theta_K$, the background models for ancestral nucleotides and all current species, and the evolutionary parameters $\alpha$, $\beta$ and $\mu_f$. We take as input the number of TFs, $K$ and the expected module length, $L$, and fix the transition probability $t = 1/L$ in $T$. Compared to the HMx model in Zhou and Wong (2004), this model has three extra free parameters, $\alpha$, $\beta$ and $\mu_f$, related to the evolutionary models. Independent Poisson priors are put on motif widths and flat Dirichlet distributions are used as priors for all the other parameters. With a given alignment for each ortholog group, we treat as missing data the locations of modules and motifs (i.e. the hidden states), the ancestral sequences and the evolutionary bonds. We develop a Gibbs sampler (called MultiModule, hereafter) to sample from the joint posterior distribution of all the unknown parameters and missing data. To consider the uncertainty in multiple alignment, we adopt an HMM-based multiple alignment [Baldi et al. (1994), Krogh et al. (1994)] conditional on the current parameter values. This is achieved by adding a Metropolis–Hastings step in the Gibbs sampler to update these alignments dynamically according to the current sampled parameters, especially the background substitution matrix $\Phi$ [equation (3)]. In summary, the input data of MultiModule are groups of orthologous sequences, and the program builds an initial alignment of each ortholog group by a standard HMM-based multiple alignment algorithm. Then each iteration of MultiModule is composed of three steps: (1) Given alignments and all the other missing data, we update motif widths and other parameters by their conditional posterior distributions; (2) Given current parameters, with probability $u$, we update the alignment of each ortholog group; (3) Given alignments and parameters, a dynamic programming approach is used to sample module and motif locations, ancestral sequences and evolutionary bonds. (See the Appendix for the details of the Gibbs sampling of MultiModule.) The probability $u$ is typically chosen in the range $[0.1, 0.3]$.

Motif and module predictions are based on their marginal posterior distributions constructed by the samples generated by MultiModule after some burn-in period (usually the first 50% of iterations). We estimate the width of each motif by its rounded posterior mean. We record the following posterior probabilities for each sequence position in all the species: (1) $P_k$, the probability that the position is within a site for motif $k$, that is, the hidden state is $M_k$ ($k = 1, 2, \ldots, K$); (2) $P_m$, the probability that the position is within a module, that is, the hidden state is $M$; (3) $P_a$, the probability that the position is aligned. All the contiguous segments with $P_k > 0.5$ are aligned (and extended if necessary) to generate predicted sites of motif $k$ given the estimated width $w_k$. The corresponding average $P_a$ over the bases of a predicted site is reported as a measure of its conservation. We collect all the contiguous regions with $P_m > 0.5$ as candidates for modules, and a module is predicted if the region contains at least two predicted motif binding sites. The boundary of a predicted module is defined by the first and last predicted binding sites it contains. We use 0.5 as the threshold for posterior probabilities. This threshold determines the trade-off between the sensitivity and the specificity of the predictions. In our experience, values in the range of $[0.5, 0.7]$ for the threshold usually give good performance for the posterior inference on the model. Smaller threshold often results in many false positive predictions.

Under the c-HMM, if we fix $r = 1 - t = 1$ in the transition matrix $T$ [equation (1)], then MultiModule reduces to a motif discovery method, assuming the existence of $K$ motifs in the sequences. This setting is useful when the motifs do not form modules, and we call it the motif mode of MultiModule.

3.2. *Multimodality and combined prediction.* Although MultiModule converges to the target posterior distribution eventually, from the examples in Section 5 we find that it usually reaches some local mode quickly and then moves around the mode for a long time. Since the waiting time for between-mode transition is exponentially long, we often run multiple short chains for MultiModule instead of one long chain, that is, we apply MultiModule to a particular data set multiple times with random initialization. In this way, it has a much greater chance to explore different major local modes which often correspond to different motif compositions of predicted modules. However, we note that our module prediction is quite consistent and that major motifs are usually predicted repeatedly in multiple runs (see Section 5.1). We employ a heuristic to select representatives of these motifs by ranking all the predicted motifs according to a Bayesian score derived in Jensen et al. (2004):

$$(4) \qquad Score = n\left[\sum_{i=1}^{w}\sum_{j} \log\left(\frac{\hat{\Theta}_{ji}}{\theta_{0j}}\right) + \log(\rho)\right] - 1.5w\log(n+3),$$

where $j = A, C, G, T$, $n$ is the number of predicted sites, $\hat{\Theta}$ is the estimated motif weight matrix constructed by predicted sites, $\theta_0$ is the background distribution, and $w$ is the width of the motif. The parameter $\rho$ ($<1$) is the prior odds of observing a motif site over a background nucleotide. We take $\rho = 1/500$, which gives good balance between the specificity of a predicted motif pattern and the number of predicted sites. Using the average $P_m$ of each sequence position over these multiple runs and the top $K$ distinct motifs ranked by (4), we define combined–predicted modules by the same criterion introduced in Section 3.1.

**4. Simulation studies.** Transcription factors Oct4, Sox2 and Nanog are believed to cooperate in the regulation of genes important to self renewal and pluripotency of embryonic stem (ES) cells [Boyer et al. (2005)]. We used the following model to simulate data sets in this study: We generated 20 hypothetical ancestral sequences, each of length 1000 bps. Twenty modules, each of 100 bps and containing one binding site of each of the three TFs, were randomly placed in these sequences. TFBSs were simulated from their known weight matrices with logo plots [Schneider and Stephens (1990)] shown in Figure 2. Then based on the choices of the background mutation rate $\mu_b$ [with $\alpha = 3\beta$ in equation (3)] and the motif mutation rate $\mu_f$, we generated sequences of three descendant species according to the evolutionary models in Section 2.3. The indel (insertion–deletion) rate was fixed to $0.1\mu_b$. After the ancestral sequences were removed, each data set finally contains 60 sequences from three species. Our simulation study was composed of two groups of data sets, and in both groups we set $\mu_f = 0.2\mu_b$ but varied

MODULE DISCOVERY IN MULTIPLE SPECIES 9

TABLE 1
*Results for the simulation study*

|  | Oct4 (60) $N_2/N_1$ | Sox2 (60) $N_2/N_1$ | Nanog (60) $N_2/N_1$ | Three motifs in total (180) | | |
|---|---|---|---|---|---|---|
|  |  |  |  | Sensitivity | Specificity | Overall |
| (A) | 38.7/57.4 | 51.6/66.0 | 40.8/46.3 | 73% | 77% | 75% |
| (B) | 27.7/45.3 | 52.2/91.3 | 27.6/37.8 | 60% | 62% | 61% |
| (C) | 22.2/36.7 | 42.4/89.6 | 23.6/39.0 | 49% | 53% | 51% |
| (A) | 18.8/24.8 | 22.7/30.6 | 21.0/33.9 | 35% | 70% | 49% |
| (B) | 9.3/29.4 | 34.0/51.1 | 21.7/31.6 | 36% | 58% | 46% |
| (C) | 5.1/8.1 | 14.2/18.2 | 8.3/14.2 | 15% | 68% | 32% |

NOTE: $N_2$ and $N_1$ refer to the numbers of correct and total predictions for each motif, respectively. TF names are followed by the numbers of true sites in parentheses. The upper and lower halves refer to the average results over 10 independently generated data sets with $\mu_b = 0.1$ and 0.4, respectively. "Overall" is the geometric average of sensitivity and specificity. For each data set, the optimal results (in terms of overall score) among three independent runs under the same parameters were used for the calculation of averages. Parameter sets (A), (B), (C) are defined as follows: (A) Module mode, $L = 100, u = 0.2$; (B) Motif mode, $u = 0.2$; (C) Motif mode, $u = 0$.

the value of $\mu_b$. In the first group, we set $\mu_b = 0.1$ to mimic the case where species are evolutionarily close. In the second group, we set $\mu_b = 0.4$ to study the situation for remotely related species. For each group we generated 10 data sets independently.

We applied MultiModule to these data sets under three different sets of program parameters: (A) Module mode, $L = 100, u = 0.2$; (B) Motif mode, $u = 0.2$; (C) Motif mode, $u = 0$. For each set of parameters, we ran MultiModule for 2,000 iterations with $K = 3$, searching both strands of the sequences. Initial alignments were built by ordinary HMM-based multiple alignment methods. If $u = 0$, these initial alignments were effectively fixed along the iterations.

The results are summarized in Table 1, which includes the sensitivity, the specificity and an overall measurement score of the performance, defined

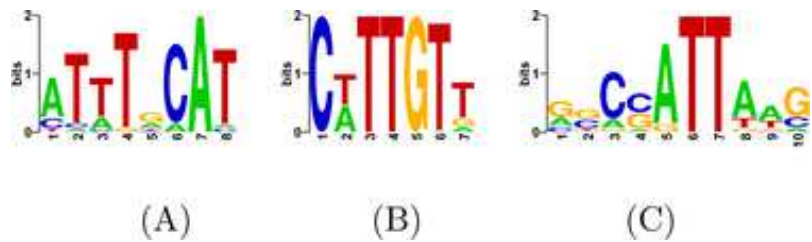

FIG. 2. *Logo plots for the motifs in the simulated studies:* (A) *Oct4,* (B) *Sox2 and* (C) *Nanog.*

TABLE 2
*Selected posterior distributions of motif widths for Oct4, Sox2, Nanog and a spurious motif*

| Width   | Oct4  | Sox2  | Nanog | Spurious |
|---------|-------|-------|-------|----------|
| 6       | 0     | 0     | 0     | 0.755    |
| 7       | 0.053 | 0.859 | 0     | 0.198    |
| 8       | 0.891 | 0.141 | 0     | 0.036    |
| 9       | 0.056 | 0     | 0.159 | 0.011    |
| 10      | 0     | 0     | 0.841 | 0        |
| [11, 15]| 0     | 0     | 0     | 0        |

NOTE: The true motif widths are 8, 7 and 10 for Oct4, Sox2 and Nanog, respectively (see Figure 2).

as the geometric average of the sensitivity and specificity. This overall score equals zero if either sensitivity or specificity is zero; it equals 1 if both of them are 1; when sensitivity = specificity = $x$, the overall score equals $x$. These properties make it a good overall measurement of predictions. One sees that updating alignments improves the performance for both $\mu_b = 0.1$ and 0.4, and the improvement is more significant for the latter setting [cf. results of (B) and (C) in Table 1]. The reason is that the uncertainty in alignments for the cases with $\mu_b = 0.4$ is higher than that for $\mu_b = 0.1$, and thus updating alignments, which aims to average over different possible alignments, has a greater positive effect. Considering module structure shows an obvious improvement for $\mu_b = 0.1$, but it is only slightly better than running the motif mode for $\mu_b = 0.4$ [cf. (A) and (B) in Table 1]. We noticed that for $\mu_b = 0.4$, MultiModule found all the three motifs under both parameter settings [(A) and (B)] for five data sets, and the predictions in (A) with an overall score of 70% definitely outperformed that in (B) with an overall score of 58%. For the other five data sets, no motifs were identified in setting (A), but in setting (B) (motif mode) subsets of the motifs were still identified for some of the data sets. We suspect that this was caused by the slower convergence of MultiModule in setting (A), because of its higher model complexity, especially when the species are farther apart. One possible quick remedy of this is to use the output from setting (B) as initial values for setting (A), which will be a much better starting point for the posterior sampling.

In all the simulation studies, motif widths were updated in the range of [6, 15] by a Metropolis–Hastings step (Section A.4). To illustrate the posterior inference of motif width in MultiModule, we report in Table 2 the histograms of the motif widths of the three TFs from a single run of 1000 Monte Carlo samples after burn-in for one of the simulated data sets with $\mu_b = 0.1$, where all the three motifs were unambiguously identified. The posterior probabilities were all concentrated on their respective true motif widths. On the other hand, we also report in this table the motif width distribution when MultiModule output a false motif (i.e., it was none of the three true motifs). One sees that, in this case, the motif width was mostly sampled as $w = 6$ and decayed very fast for $w > 6$. This was due to the fact that we restricted the motif width to be between 6 and 15. If we removed this restriction, the width would further decrease to smaller values, which would be a good indication that this motif might be spurious.

**5. Applications to biological data sets.** We tested MultiModule on two annotated data sets from mammals and fruit flies. Our computational predictions were compared to experimental validations reported previously. Detailed comparison with several published motif and module discovery methods was conducted based on these data sets. Hereafter we say that a predicted site is a correct prediction or that a predicted site overlaps an experimentally verified site if the starting position of the predicted site is within 3 bps to that of a verified site. This definition is used for assessing the performance of all the computational methods mentioned in this article. In the following examples, we set $u = 0.2$ to update ortholog alignments in MultiModule.

5.1. *Muscle-specific genes in mammals.* Our first test data set is the 24 skeleton–muscle-specific genes of human and mouse orthologs [Wasserman et al. (2000)]. We combined putative dog orthologs based on UCSC genome alignment (http://www.genome.ucsc.edu). The muscle-specific expression of these genes is controlled by five TFs, MEF, MYF, SP1, SRF and TEF, with 16, 25, 21, 14 and 7 experimentally validated binding sites in the human genes, respectively [Thompson et al. (2004)]. These binding sites form 24 validated modules. Here a validated module is defined as a sequence fragment containing at least two TFBSs satisfying the condition that the distance between any two neighboring sites is $< 100$ bps. The module boundaries are defined by the first and last TFBSs it contains. The total length of these modules is 2716 bps, distributed in 21 human genes. Upstream 3 kb of the human, mouse and dog orthologs were extracted and aligned by mLAGAN [Brudno et al. (2003)]. Based on these alignments, we calculated a conservation score ($CS$) for each sequence position. Using a threshold, we N-masked nonconserved bases which are more than 1 kb from the TSSs (transcription start site), that is, all the bases within 1 kb are kept irrespective of their $CS$s. The purpose is to detect promoter and conserved distal enhancer elements. Repeats were masked by "N" using a repeat-masking program (http://repeatmasker.org/). The preprocessing reduced our data set to an average effective length ("N"s are not counted) of 1604 bps per sequence.

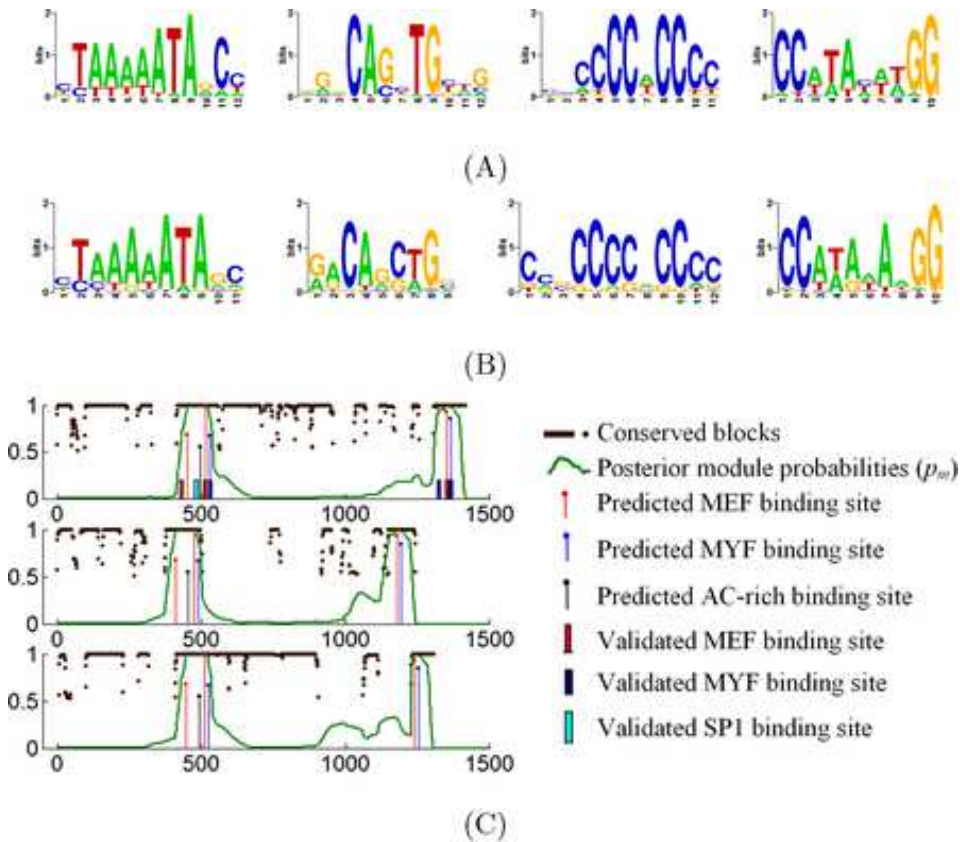

FIG. 3. *Motif and module predictions for the muscle-specific gene set. Logo plots for* (A) *experimentally validated and* (B) de novo *predicted binding sites for MEF, MYF, SP1 and SRF in the human genes.* (C) *The Bayesian inference for the gene TNNI1 in the human (top), mouse (middle) and dog (bottom) orthologs. The green curved lines report the posterior module probabilities ($P_m$). Conserved regions (defined as posterior alignment probability $P_a > 0.5$) are indicated by brown bars and dots, of which the vertical heights report the corresponding values of $P_a$. Predicted motif sites and experimentally validated sites are indicated by vertical lines and bars, respectively.*

We applied MultiModule with $K = 5$ and $L$ (expected module length) = 200 bps. Our pilot study suggests that running multiple short chains is more efficient for MultiModule (see supplemental notes). Thus, we ran the program 50 times independently and 1000 iterations each run. All the predicted motifs were ranked by their Bayesian scores [equation (4)], and the top five distinct motifs corresponded to the binding patterns of SRF, MEF, SP1 and MYF [Figure 3(A) and (B)], and an AC-rich motif which seems to be a repetitive pattern. We note that MultiModule failed to discover any motifs close to that of TEF, mainly due to the fact that the TEF sites are not enriched enough in this data set (only seven sites in 24 sequences). For the other four known TFs, MultiModule predicted a total of 97 sites in the human genes from the top five ranking motifs, and 45 of them overlap corresponding validated sites. Thus, it achieves a sensitivity of 59% and a specificity of 46% for these four motifs (Table 3). Note that the specificity is likely to be underestimated since our predictions may include some functional binding sites which have not been experimentally validated yet. The 50 runs contained at least two distinct modes in module composition, denoted by mode A and mode B. In mode A, MultiModule found five motifs corresponding to MEF, MYF, SP1, SRF and the AC-rich motif. In mode B, MultiModule found four motifs, including MYF, SP1, MEF and the AC-rich motif, with some validated SRF sites contained in the predicted MEF sites, that is, one motif was a mixture of MEF and SRF. One sees that these two motifs are both AT-rich in the middle [Figure 3(A)]. Please note that it is possible that MultiModule output fewer motifs than the pre-specified $K$, when the posterior motif site probabilities $P_k$ of all sequence positions are $<0.5$ (the posterior probability threshold) for some $k$. We randomly selected one representative from each mode and checked their overlaps in base pair level with validated modules. Both modes predicted modules at a sensitivity of 65% and a specificity of 41% approximately (Table 4). We averaged posterior module probabilities over the 50 independent runs for each sequence position and used these average $P_m$'s to generate our combined predictions with the top five predicted motifs. This combined prediction shows a significant improvement in specificity without loss of much sensitivity (Table 4). The Bayesian inference by marginal posterior probabilities is illustrated in Figure 3(C) using the gene TNNI1 as an example, which contains two known modules. One sees that (1) high peaks of $P_m$ emerge at the two known modules; (2) the posterior module and motif probabilities are coupled in conserved regions and thus show similar shape among the orthologs.

We note that the bases in the combined-predicted modules showed a very high average $P_m$ and the $P_m$ values for 82% of them were $>0.9$ [Figure 4(A)]. This implies that the module predictions were quite consistent among independent runs despite the slightly different motif composition. The average $P_a$'s of predicted motifs and modules (Tables 3 and 4) were much higher than the overall average of all the sequence positions (58%), which indicates that functional elements are more likely to be located in aligned blocks. We observed that the background mutation rate $\mu_b$ was significantly higher than that of TFBSs $\mu_f$ [Figure 4(B)]. From the respective definitions of $\mu_b$ and $\mu_f$, one sees that, for a TFBS, the equivalent mutation rate comparable to the definition of $\mu_b$ is $<\mu_f$. Thus, the observation that $\mu_b > \mu_f$ convinces that even within aligned blocks, TFBSs still show a significantly lower evolutionary rate than their surrounding background nucleotides.

TABLE 3
*Predicted motifs in the muscle-specific genes*

| TFs | Human Validated | Human Predicted | Human Overlaps | Mouse Predicted | Dog Predicted | Avg. $P_a$ (%) |
|-----|-----------|-----------|----------|-----------------|---------------|----------------|
| MEF | 16 | 25 (10*) | 13 (7*) | 26 | 24 | 99 |
| MYF | 25 | 34 | 13 | 35 | 31 | 97 |
| SP1 | 21 | 16 | 8 | 21 | 16 | 70 |
| SRF | 14 | 22 | 11 | 22 | 22 | 98 |
| Total | 76 | 97 | 45 | 104 | 93 | 93 |

NOTE: Tabulated are the numbers of validated TFBSs, predicted sites and overlapping sites (i.e., correct predictions) for the human sequences. We also include the total number of predicted sites in the mouse and dog sequences.
*Among the predicted MEF sites, 10 of them overlap the predicted SRF sites, in which seven turn out to be experimentally validated SRF binding sites.

5.2. *Early developmental genes in Drosophila.* Regulatory regions that control early body development in Drosophila melanogaster (Dm, hereafter) were identified in previous experimental studies [e.g., Berman et al. (2002)]. As a test of MultiModule, we extracted all the identified regulatory regions that interact with at least one of the three TFs, Bicoid (Bcd), Hunchback (Hb) and Krüppel (Kr), which form complex combinatorial patterns that regulate early developmental genes. These extracted regions form a data set of 26 Dm sequences. We further extracted orthologous regions in Drosophila pseudoobscura (Dp, hereafter) based on UCSC genome alignment and obtained 23 of them. Thus, our full data set contains 49 sequences with an average length of 1209 bps. In this data set, 34 functional binding sites in Dm with experimental validations are available based on TRANSFAC 9.1 release [Wingender et al. (2000)]: 12 Bcd sites in three sequences, 14 Hb sites

TABLE 4
*Predicted modules in the muscle-specific genes*

|  | Predicted modules (bps) | Overlaps (bps) | Sensitivity (%) | Specificity (%) | Avg. $P_a$ (%) |
|---|---|---|---|---|---|
| Mode A | 4159 | 1728 | 64 | 42 | 85 |
| Mode B | 4412 | 1768 | 65 | 40 | 87 |
| Combined | 2835 | 1566 | 58 | 55 | 89 |

NOTE: Tabulated are the total numbers of bases in the predicted modules in the human gene set with validated modules (the 21 human genes with at least one defined module of validated TFBSs). Overlaps report the number of overlaps between the predicted and validated modules in base pair level. Sensitivity = (# overlaps) / (# bases in all validated modules = 2716 bps). Specificity = (# overlaps) / (# bases in the predicted modules).

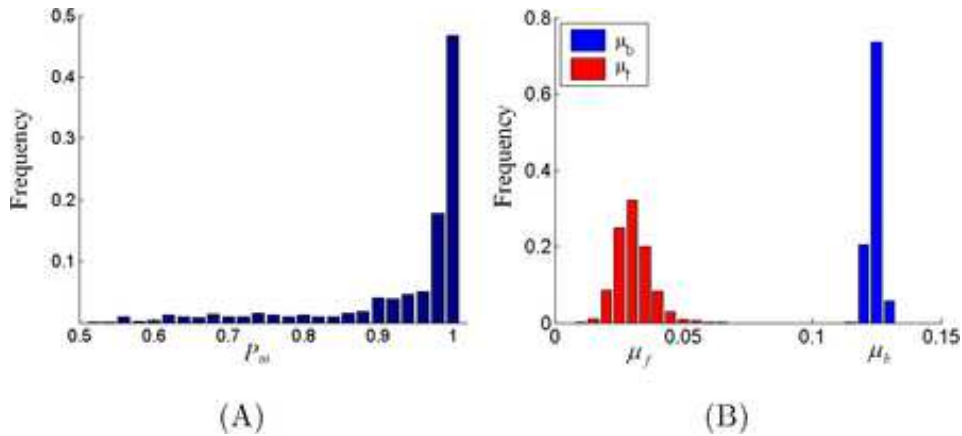

FIG. 4. (A) *Histogram of the average posterior module probabilities ($P_m$) over the 50 independent runs of all the bases in the combined-predicted modules.* (B) *Posterior distributions of the background and motif mutation rates, $\mu_b$ and $\mu_f$, respectively, calculated by pooled samples from the 50 runs.*

in four sequences, and 8 Kr sites in two sequences. The genes with validated sites are referred to as validated gene set in this example.

We applied MultiModule to this data set with $K=3$ and $L=200$ for 50 independent runs, each 1000 iterations. We ranked all the predicted motifs in multiple runs by the same Bayesian score [equation (4)], and the top two distinct motifs correspond to the known Hb and Kr binding patterns [Figure 5(A) and (B)]. The third motif resembles the known Bcd binding pattern [Figure 5(C)]. Our predicted motif binding sites overlap substantially with validated ones for the three TFs with an overall sensitivity of 44% and specificity of 47% (Table 5). Some of our predicted Kr sites turned out to be validated Bcd sites, which is consistent with the fact that these two TFs actually bind to some overlapping functional sites. Such sites are not counted as correct predictions for Bcd in Table 5. Since the majority of the sequences in this data set are not annotated for TFBSs, we include in Table 5 the sum of squared distances (SSD) between the predicted and the known motif weight matrices as another quality measure of the predictions. These SSDs are approximately 1/10 of the expected SSDs between a random weight matrix and the known ones for the three TFs.

We combined all the 50 independent runs to generate our combined predictions. In this way, we predicted 70 modules covering 16855 bps (Table 6): 86%, 49% and 30% of these modules contain the predicted Hb, Kr and Bcd binding sites, respectively. We used these combined predictions to define regulatory interactions among the maternal gene Bcd and the four gap genes in our data set, Hb, Kr, Gt and Kni, which are the most important

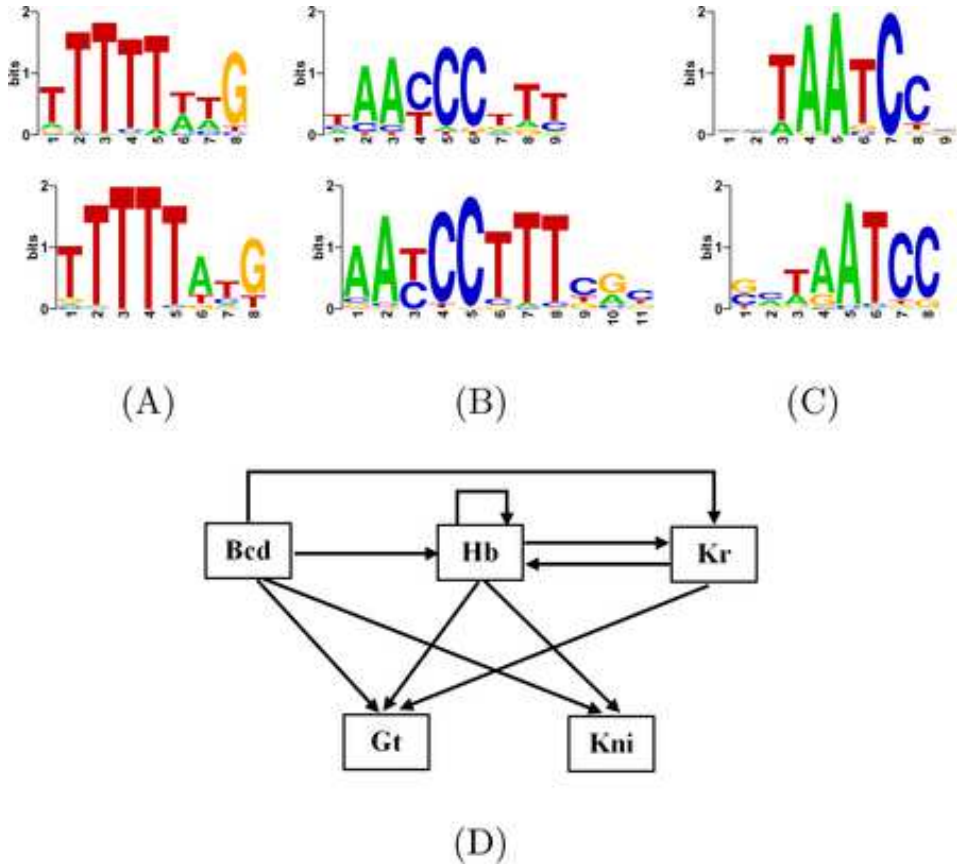

FIG. 5. *Results of the Drosophila early developmental genes. Logo plots for the known (upper panel) and the predicted (lower panel) motifs in Dm for the three TFs:* (A) *Hb,* (B) *Kr and* (C) *Bcd. The predicted motifs in Dp of these three TFs are identical to those in Dm.* (D) *The predicted regulatory network of the genes Bcd, Hb, Kr, Gt and Kni in early Drosophila body patterning. An arrow from gene Y to gene X indicates that gene Y regulates gene X based on our predicted modules.*

TFs controlling early body patterning in Drosophila. Known regulatory interactions among these genes are available based on previously reported experiments [Sanchez and Thieffry (2001)]. In our simplified analysis, gene X is defined to be regulated by gene Y if the regulatory region of gene X contains a predicted module composed of at least one binding site for the TF encoded by gene Y. Remarkably, all our predicted interactions [Figure 5(D)] are exactly identical to those known ones, and our analysis recovers all the known interactions with Bcd, Hb and Kr as regulators. In this data set, the predicted motifs and modules also showed higher conservation (Tables 5 and 6) compared with the overall average $P_a$ of 33%.

TABLE 5
*Motifs predicted in the Drosophila early developmental genes*

| TFs | Validated genes in Dm | | | Full Dm Predicted (SSD) | Full Dp Predicted (SSD) | $P_a$ (%) |
|---|---|---|---|---|---|---|
| | Validated | Predicted | Overlaps | | | |
| Hb | 14 | 18 | 9 | 156 (0.31) | 143 (0.28) | 57 |
| Kr | 8 | 4 | 4 | 54 (0.41) | 37 (0.38) | 53 |
| Bcd | 12 | 10 | 2 | 65 (0.36) | 60 (0.41) | 75 |
| Total | 34 | 32 | 15 | 275 | 240 | 56 |

NOTE: Predictions in validated gene set and in full gene set are tabulated. SSD is the sum of squared distance between a predicted and the corresponding known weight matrices. Other fields are defined similarly to those in Table 3. Validated genes in Dm refer to the genes with known TFBSs for the corresponding TFs. Exact binding sites are not available for the remaining genes although they are known to be controlled by these TFs.

5.3. *Comparing with other methods.* We compared the performance of MultiModule on the two data sets with that of AlignACE [Roth et al. (1998)], CompareProspector [Liu et al. (2004)], EMnEM [Moses, Chiang and Eisen (2004)] and CisModule [Zhou and Wong (2004)]. AlignACE finds multiple motifs using a repeatedly masking strategy. We ran the program under its default setting to search for motifs with different combinations of input parameters for motif width and expected number of sites. For these two data sets, AlignACE output 40 to 60 motifs for each run, and we repeated the program five times for each data set independently, which generated around 250 motifs. CompareProspector searches for motifs in one species with a given conservation score for each sequence position calculated based on a multiple alignment. We ran CompareProspector with each known motif width and a total of 300 independent runs for each data set. EMnEM takes as input one alignment for each ortholog group with a given phylogenetic tree. We input the alignments built by mLAGAN [Brudno et al. (2003)] and used a phylogenetic tree with branch length (in the unit of substitution per

TABLE 6
*Predicted modules in the Drosophila early developmental genes*

| TFs | Modules (#) | Modules (bps) | $P_a$ (%) | Motifs in % of modules | | |
|---|---|---|---|---|---|---|
| | | | | Hb (%) | Kr (%) | Bcd (%) |
| Dm | 37 | 8650 | 42 | 84 | 54 | 30 |
| Dp | 33 | 8205 | 45 | 88 | 42 | 30 |
| Total | 70 | 16855 | 43 | 86 | 49 | 30 |

NOTE: Tabulated are the number and total length (in the unit of base pair) of the predicted modules, their average alignment probability ($P_a$) and the percentage of the modules containing each detected motif.

site) estimated by Xie et al. (2005) from mammalian upstream sequences for the first data set. We ran EMnEM with each known motif width $w$ and set all the $w$-mers as initial consensus. For the above three methods, we defined representative output for each known motif by the highest score predicted motif (as reported by respective programs) that match the known pattern. CisModule is a single-species module discovery method. We ran it 50 times independently under the same parameters ($K$ and $L$) as those used in MultiModule. The predicted motifs of CisModule were ranked by the same score function [equation (4)]. For all the methods, we used exactly the same sequences after preprocessing as described in the previous sections.

The two test data sets in total contain 117 validated TFBSs for the eight known motifs. Table 7 summarizes the performance of each method on these data sets with respect to motif identification and site prediction. Multi-Module identified more known motifs than all the other methods. It also found much more validated TFBSs than the other methods did, with at least about 70% of improvement in sensitivity: MultiModule detected 60 validated TFBSs, while the other methods detected at most 35 validated sites. In addition, the specificity of MultiModule is the highest one among all the programs for these tests. This indicates that the high sensitivity of MultiModule does not come at the expense of specificity. In terms of overall performance, MultiModule shows 53% to 81% of improvement compared to the other four methods.

TABLE 7
*Performance comparison on the two test data sets*

| Methods | # correctly identified motifs | For correctly identified motifs | | | | |
|---|---|---|---|---|---|---|
| | | # of predicted | # of overlaps | Sensitivity (%) | Specificity (%) | Overall (%) |
| ALnACE | 6 | 106 | 31 | 26 | 29 | 28 |
| CompPr | 6 | 64 | 27 | 23 | 42 | 31 |
| EMnEM | 5 | 102 | 35 | 30 | 34 | 32 |
| CisMod | 5 | 110 | 35 | 30 | 32 | 31 |
| MltMod | 7 | 129 | 60 | 51 | 47 | 49 |

NOTE: We compare different methods based on their predictions with respect to the validated gene sets (i.e., genes with validated TFBSs) which are exactly the same as those used in the applications of MultiModule to these data sets. "ALnACE," "CompPr," "EMnEM," "CisMod" and "MltMod" refer to AlignACE, CompareProspector, EMnEM, CisModule and MultiModule, respectively. We report the number of known motifs each method identified. For correctly identified motifs, the numbers of predicted sites and correct predictions (# overlaps) are reported. Sensitivity is calculated by (# overlaps) / (total number of validated TFBSs = 117). Specificity is calculated by (# overlaps) / (# predicted sites). Overall score is defined as the geometric average of sensitivity and specificity.

**6. Discussion.** We have proposed and illustrated a new computational approach based on a coupled hidden Markov model for *de novo* discovery of CRMs and motifs in sequences from multiple species. Our simulation and test results convey three pieces of information about this approach. First, modeling sequence orthology provides more information than using a conservation score or simply pooling sequences from multiple species into a heterogeneous data set as illustrated from the comparison with other methods. Second, the use of module structure to identify clusters of motif patterns is usually more powerful than identifying each motif independently. Third, updating multiple alignments improves the sensitivity of motif prediction. From this study, we observe that for species within mammals or within Drosophila, aligned regions are usually much longer than the width of TFBSs, and TFBSs definitely show lower mutation rates compared with aligned background nucleotides as shown in Figure 4(B), which is also true for the Drosophila data set.

Since MultiModule samples from a complicated joint probability distribution by a Gibbs sampler, the problem of multimodality needs to be considered. We observe that it is helpful to integrate out weight matrices and other parameters even in an approximate sense (Section A.6) for the convergence of MultiModule. To further alleviate the possibility of local traps, we combine samples from multiple randomly initialized chains to construct superior estimates in practice. An alternative approach to this end is the use of more sophisticated Monte Carlo algorithms to handle multimodality, such as the parallel tempering [Geyer (1991)] or the equi-energy sampler [Kou, Zhou and Wong (2006)]. The computational complexity of MultiModule is approximately proportional to $2^N KL$, where $N$ is the number of species, $K$ is the number of TFs, and $L$ is the total length of sequences, which is scalable with a reasonable selection of orthologous species. It is worth mentioning that the complexity of the motif mode of MultiModule is linear in $N$. The use of c-HMM is not restricted to *de novo* discovery. With given PWMs and other parameters, MultiModule can be used to scan for modules in ortholog groups. From our experience, for *de novo* motif finding, MultiModule is suitable to handle sequences with an average length less than 2 kb. If the average search region is much larger, some preprocessing is needed to reduce it to save the computational cost and to increase the motif site enrichment. This was the reason why we removed the nonconserved bases in the first application of the muscle-specific genes.

For the current implementation of MultiModule, we assume that the number of motifs $K$ is fixed and known. But in real module discovery applications, the exact value of $K$ is usually unknown. It would be desirable to develop a coherent approach to estimate this number simultaneously. However, up to now, we do not have an efficient method to complete this task. The dimensionality of the parameter space is determined by $K$. With all kinds of missing data in our model, it is impossible to integrate out both the parameters and the missing data to obtain the marginal distribution of $K$. Thus, we decide to fix $K$ for the current implementation. When $K$ is set to be smaller than the true number of motifs, the program usually finds subsets of the motifs that are highly enriched (some repetitive patterns may be included as well, such as the AC rich pattern in the first application). When $K$ is set to be greater than the true value, the program may output fewer than $K$ motifs, as we pointed out in Section 5.1. Our suggestion is to apply some pilot runs of MultiModule with different $K$ in a reasonable range, say, between 2 and 6, and check how many distinct motifs it actually outputs. Then one may make a decision on the suitable value of $K$ and perform multiple runs to generate final predictions. In our experience, this ad hoc approach is usually acceptable for most applications.

In this article the module structure is modeled by a one-step Markov chain with two states, which specifies a geometric distribution on the length of a module. This is definitely a simple approximation to the biological reality, which only captures the co-localization of multiple motif sites, that is, the model puts some constraints on the possible locations of the motif sites in the same module. Our current approach should be viewed as a very first step to incorporate module and phylogenetic structures into *de novo* motif discovery. There exists substantial room for further development in the direction. One may extend the HMM used in this article to a higher-order Markov chain such that the transition probability to a motif depends on the previous motif in the module. This allows the model to estimate possible synergistic interactions between neighboring motifs. Other refinement of the model, such as more comprehensive phylogenetic tree topology and more sophisticated background model, is expected to enhance the utility of this approach. However, with the increasing model complexity, the computational efficiency and robustness of the statistical inference based on posterior sampling will be more challenging.

## APPENDIX: THE GIBBS SAMPLER FOR THE C-HMM

Suppose the data set of interest contains sequences of $n$ genes from $N$ species, $S_i^{(m)}$ ($i = 1, 2, \ldots, n, m = 1, 2, \ldots, N$). Without loss of generality, we assume $n = 1$, since the sampling procedure for the $i$th ortholog group is similar for all $i$. Thus, we simply denote the input sequences as $S = [S^{(1)}, \ldots, S^{(N)}]$. Define $\Theta = [\theta_0, \Theta_1, \ldots, \Theta_K]$, where $\theta_0 = [\theta_0^{(0)}, \theta_0^{(1)}, \ldots, \theta_0^{(N)}]$ denotes the background distributions for the ancestral sequence ($\theta_0^{(0)}$) and the current species ($\theta_0^{(m)}; m = 1, 2, \ldots, N$), and $\Theta_k$ is the PWM for the $k$th motif ($k = 1, 2, \ldots, K$). Each background model is an i.i.d. multinomial distribution. We estimate $\theta_0^{(m)}$ for the current species ($m = 1, 2, \ldots, N$) before the Gibbs sampling iteration, and thus effectively assume that they are given. Let $W = [w_0, w_1, \ldots, w_K]$ be the widths of the background model ($w_0 = 1$) and the motifs (i.e., $\Theta_k$ is a $4 \times w_k$ matrix). The transition probability matrix $T$ is defined in (1), and we denote $q = [q_0, q_1, \ldots, q_K]$. The neutral evolution of background nucleotides is characterized by the parameter vector $\phi_b = [1 - \mu_b, \alpha, 2\beta]$ [equation (3)]. The probability of breaking an evolutionary bond between any base of an aligned TFBS and that of its ancestral site is $\mu_f$. Denote the hidden states by $Y$, which indicate whether the observed nucleotides are located in a background ($B$) or module ($M$) region. For those in a module, the hidden states $Y$ also specify whether they are within-module background ($M_0$) or motif sites ($M_1$ to $M_K$). Thus, the hidden states imply the locations of modules and motif sites. We further use $A$ to denote the multiple alignment of $S^{(1)}, \ldots, S^{(N)}$, $Z$ to denote the ancestral sequence and $V$ to denote the evolutionary bonds of aligned TFBSs. Conceptually, we treat $A, Y, Z$ and $V$ as missing data and denote them by $D_{mis} = [A, Y, Z, V]$. All the parameters are denoted by $\Psi = [W, \Theta, T, q, \phi_b, \mu_f]$.

**A.1. Prior and posterior distributions.** MultiModule takes expected module length $L$ and the number of motifs (TFs) $K$ as input and fixes the transition probability from a module state to a background state $t = 1/L$. We put independent Poisson priors with mean $\lambda = 10$ for motif widths. Flat Dirichlet (Beta) priors are prescribed to all the other parameters. More specifically, we use a flat product Dirichlet of dimension $4 \times w_k$ as the prior distribution for $\Theta_k$ ($k = 1, 2, \ldots, K$) (i.e., the parameter for this product Dirichlet distribution is a $4 \times w_k$ matrix with all elements $= 1$). We put four-, $(K+1)$- and three-dimensional flat Dirichlet priors on $\theta_0^{(0)}$, $q$ and $\phi_b$, respectively. The prior distributions for $r$ and $\mu_f$ are both Beta(1, 1). With these prior distributions specified, one can write down the joint posterior distribution of all the parameters and missing data,

$$(5) \qquad P(\Psi, D_{\text{mis}}|S) \propto P(D_{\text{mis}}, S|\Psi)\pi(\Psi),$$

where $\pi(\Psi)$ denotes the joint prior distribution. Of interest are the locations of motif sites and modules, that is, the hidden states $Y$. One wants to perform inference based on the marginal posterior distribution of $Y$ given all the sequence data $S$,

$$(6) \qquad P(Y|S) \propto \int_\Psi \sum_{A,Z,V} P(Y, A, Z, V, S|\Psi)\pi(\Psi)\,d\Psi,$$

where all the other unknown variables are marginalized out. However, the integral in (6) has no analytical solution, and thus, we devise a Gibbs sampling approach to generate samples from the joint distribution [equation (5)]. Based on these samples, one can easily construct empirical marginal posterior distributions of interesting variables ($Y$ in this case).

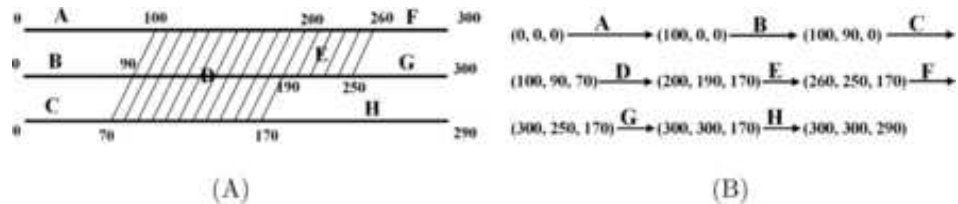

FIG. 6. *The map from the alignment of three sequences in* (A) *to its path representation in 3-D space in* (B). *The path starts from* (0, 0, 0), *moves along A-B-C-D-E-F-G-H, and ends at* (300, 300, 290). *Shaded areas in* (A) *represent aligned blocks. Coordinates of aligned positions increase simultaneously along the path, such as those of sub-path D from* (100, 90, 70) *to* (200, 190, 170).

Our Gibbs sampler contains three steps: (1) sample $A$ given $\Psi$; (2) sample $[Y, Z, V]$ jointly given $A$ and $\Psi$; (3) sample $\Psi$ given $[Y, Z, V]$ and $A$. To simplify the description, we first discuss how to sample from the conditional distributions assuming that the alignment $A$ is given. Then we discuss how to update $A$ by one more conditional sampling step by a Metropolis–Hastings algorithm. We also describe how to integrate out all the parameters $\Psi$ in (5) in an approximate manner, so that we are effectively sampling from $P(D_{mis}|S)$. We find this implementation of a collapsed Gibbs sampler [Liu (1994)] improves the convergence of MultiModule.

**A.2. Path representation of the c-HMM.** Any multiple alignment of the orthologs $S = [S^{(1)}, \ldots, S^{(N)}]$ can be viewed as a path in the $N$-dimensional space from $(0, 0, \ldots, 0)$ to $(L_1, L_2, \ldots, L_N)$, with $L_m$ the length of $S^{(m)}(m = 1, 2, \ldots, N)$. The path is composed of a series of $N$-dimensional points, and the coordinates of each point are the last visited positions of these $N$ sequences. Figure 6 shows the map from a multiple sequence alignment to a path in 3-D space, which visits all the sequence positions in a unique order. This is a natural generalization of the 2-D path representation of a pair-wise alignment.

Suppose the current alignment path $A = a_1 a_2 \cdots a_{L(A)}$, where $L(A)$ is the total number of points and $a_d$ is the $d$th point in this path $[d = 1, 2, \ldots, L(A)]$. Denote the coordinates of $a_d$ by $C(d) = [c_d^{(1)}, \ldots, c_d^{(N)}]$, especially, $C(1) = [0, \ldots, 0]$ and $C(L(A)) = [L_1, \ldots, L_N]$. Define the emission components ($EC$) of $a_d$ as the subset of its $N$ coordinates which increase compared with the corresponding coordinates of $a_{d-1}$, that is, $EC(d) = \{m | c_d^{(m)} > c_{d-1}^{(m)}, m = 1, \ldots, N\}$. For instance, in Figure 6, the emission component of a point in sub-path A is the single element set $\{1\}$, since along this sub-path only the first sequence emits nucleotides, while the emission components of a point in sub-path D is $\{1, 2, 3\}$. We further denote the nucleotides emitted at $a_d$ by $X(d) = \{x_d^{(m)} | m \in EC(d)\}$, where $x_d^{(m)}$ is the nucleotide at position $c_d^{(m)}$ in $S^{(m)}$. Let $|EC(d)|$ denote the size of the emission components of $a_d$, that is, $X(d)$ contains nucleotides from $|EC(d)|$ species. If $|EC(d)| \geq 2$, the emission components are aligned to each other at $a_d$. Let $Y(d) = [y_d^{(1)}, \ldots, y_d^{(N)}]$ be the hidden states of $a_d$, with $y_d^{(m)}$ the hidden state at $c_d^{(m)}$ ($m = 1, 2, \ldots, N$). The following constraints are implied by the definition of the c-HMM:

(i) $y_d^{(i)} = y_d^{(j)}$, if $x_d^{(i)}$ and $x_d^{(j)}$ are aligned ($i, j = 1, 2, \ldots, N$);
(ii) $y_d^{(m)} = y_{d-1}^{(m)}$, if $m \notin EC(d)$;
(iii) $y_d^{(m)} = y_{d+1}^{(m)} = \cdots = y_{d+w_k-1}^{(m)} = M_k$, if $[c_d^{(m)}, c_{d+w_k-1}^{(m)}]$ is a binding site of motif $k$ in $S^{(m)}$.

Constraint (i) says that any aligned nucleotides share the same hidden state (the coupled state). Constraint (ii) is due to the fact that if $m \notin EC(d)$, which implies that $c_d^{(m)} = c_{d-1}^{(m)}$, then $x_d^{(m)}$ and $x_{d-1}^{(m)}$ actually refer to the same nucleotide in $S^{(m)}$ and thus have the same hidden state. For example, in Figure 6, the third coordinates of the points in sub-path E are identical, and they all refer to position 170 of $S^{(3)}$. Constraint (iii) is consistent with the definition that a motif binding site is treated as one state ($M_k$) in our model. We define change points of the path by

$$CP(A) = \{a_d \mid C(d+1) - C(d) \neq C(d) - C(d-1), 1 < d < L(A)\},$$

which are the points where the alignment path changes its direction in the $N$-dimensional space. For instance, all the points shown in Figure 6(B) are change points except the starting and ending points. By this path representation of an alignment, we effectively augment the hidden state to an $N$-dimensional vector such that the model reduces to a Markov chain in the augmented state space. Then one may use the Markovian property to derive recursive algorithms to sample the hidden states exactly given parameter values and an alignment.

**A.3. Sample $[Y, Z, V]$ given $\Psi$ and $A$.** Denote $Y[i, j] = [Y(i), Y(i+1), \ldots, Y(j)]$ and $X[i, j] = [X(i), X(i+1), \ldots, X(j)]$ for any two points $a_i$ and $a_j$ ($j \geq i$) in the path, especially, $Y[i, i] = Y(i)$ and $X[i, i] = X(i)$. Define

(7) $$f_d(y) = P(Y(d) = y, X[1, d] | \Psi, A),$$

for $d = 1, 2, \ldots, L(A)$, where $y$ is an $N$-dimensional vector in the augmented state space. Let $y_H = \{[y^{(1)}, \ldots, y^{(N)}] \mid y^{(m)} = H, \forall m \in EC(d)\}$ for $H = B$ (background state) or $M$ (module state), which represents an $N$-dimensional state vector with $H$ as its components in $EC(d)$. Then we have the following recursive summations to calculate (7):

$$f_d(y_B) = P(X(d)|\theta_0, \phi_b) \sum_y Tr(y_B|y, r, t) f_{d-1}(y), \tag{8}$$

$$f_d(y_M) = \sum_{k=0}^{K} \bigg[ q_k P(X[d-w_k+1, d]|\Theta_k, \mu_f, \phi_b)$$

$$\times \sum_y Tr(y_M|y, r, t) f_{d-w_k}(y) \bigg], \tag{9}$$

where $\mathrm{Tr}(y_H|y, r, t)$ specifies the transition probability from $y$ to $y_H$, and $P(X(d)|\theta_0, \phi_b)$ and $P(X[d-w_k+1,d]|\Theta_k, \mu_f, \phi_b)$ are the marginal probabilities of emitting $X(d)$ and $X[d-w_k+1, d]$, given that the current state is $B$ and $M_k$ ($k = 0, 1, \ldots, K$), respectively. Recall that $M_k$ is the decomposed state of $M$, which indicates whether the current hidden state is within-module background ($M_0$) or one of the $K$ motifs ($M_1$ to $M_K$).

Since a TFBS is treated as one state as a whole, no change points are allowed in the interval $[d-w_k+1, d]$ for $k = 1, \ldots, K$ in the summation of (9). In other words, motif sites are not allowed to be located across any change points of the alignment. We calculate the transition probabilities in (8) and (9) according to the definition in (2),

$$Tr(y_H|y, r, t) = \frac{1}{|EC(d)|} \sum_{m \in EC(d)} T(y^{(m)}, H), \quad \text{for } H = B \text{ or } M, \tag{10}$$

where $y = [y^{(1)}, \ldots, y^{(N)}]$ and $y^{(j)} = y_H^{(j)}$ for $j \notin EC(d)$ due to constraint (ii) described in the previous section. If $a_{d-w_k}$ is not a change point ($k \in \{0, 1, \ldots, K\}$), that is, $EC(d) = EC(d-i)$, $i = 1, 2, \ldots, w_k$, equation (10) can be simplified as $Tr(y_H|y, r, t) = T(H', H)$, where $y^{(m)} = H'$ for $m \in EC(d)$. For example, consider a point $C(d) = [250, 240, 170]$ in sub-path E in Figure 6, whose $EC(d) = \{1, 2\}$. Suppose its hidden state is $Y(d) = [B, B, M]$. Since the previous point $C(d-1) = [249, 239, 170]$ is in the same sub-path (not a change point), we know that the hidden state $Y(d-1)$ is of the form $[y, y, M]$ with $y = B$ or $M$, and

$$Tr(Y(d) = [B, B, M]|Y(d-1) = [y, y, M], r, t) = T(y, B).$$

If $Y(d) = [M_k, M_k, M]$ ($k = 0, 1, \ldots, K$), we will consider the point $(d - w_k)$ in a similar way given that $(d - w_k) \in$ sub-path E.

If $|EC(d)| \geq 2$, the current point $a_d$ emits a group of aligned nucleotides. In order to calculate the marginal emission probabilities, one needs to sum over all possible ancestral nucleotides. More specifically,

$$P(X(d)|\theta_0, \phi_b) = \sum_z \bigg[ \theta_0^{(0)}(z) \prod_{m \in EC(d)} \Phi_b(z, x_d^{(m)}) \bigg], \tag{11}$$

where $\theta_0^{(0)}(z)$ is the probability of observing nucleotide $z$ under the ancestral background model, and $\Phi_b$ is the neutral substitution matrix defined in (3). For $k = 1, 2, \ldots, K$,

$$
\begin{aligned}
&P(X[d - w_k + 1, d] | \Theta_k, \mu_f, \phi_b) \\
&\quad = \prod_{i=1}^{w_k} \left[ \sum_{z_i} \Theta_{ki}(z_i) \prod_{m \in EC(d)} P(x_{d-w_k+i}^{(m)} | z_i, \mu_f, \Theta_{ki}) \right],
\end{aligned}
\tag{12}
$$

where $\Theta_{ki}$ is the weight vector at the $i$th position of motif $k$, that is, the $i$th column of $\Theta_k$, and $P(x_{d-w_k+i}^{(m)} | z_i, \mu_f, \Theta_{ki})$ is the probability that the ancestral nucleotide $z_i$ evolves to the descendant nucleotide $x_{d-w_k+i}^{(m)}$ under the evolutionary model for TFBSs,

$$
\begin{aligned}
&P(x_{d-w_k+i}^{(m)} | z_i, \mu_f, \Theta_{ki}) \\
&\quad = (1 - \mu_f) \cdot \mathbf{1}(x_{d-w_k+i}^{(m)} = z_i) + \mu_f \Theta_{ki}(x_{d-w_k+i}^{(m)}).
\end{aligned}
\tag{13}
$$

For $k = 0$, equation (12) reduces to equation (11) for within-module background. If $|EC(d)| = 1$, the calculation reduces to single-species situations, such as in CisModule [Zhou and Wong (2004)].

Using the recursions of equations (8) and (9) along the alignment path $A$ for $d = 1, 2, \ldots, L(A)$, one can calculate the marginal probability of observing all the sequences $S = [S^{(1)}, \ldots, S^{(N)}]$ given the current parameters and alignment, that is,

$$
P(S | \Psi, A) = P(X[1, L(A)] | \Psi, A) = \sum_y f_L(y),
\tag{14}
$$

where all the hidden states $Y$, ancestral nucleotides $Z$ and evolutionary bounds $V$ are summed over.

Based on the partial summations calculated in equations (8) and (9), we sample $[Y, Z, V]$ sequentially from $d = L(A)$ to $d = 1$. Suppose we have sampled the hidden state of $a_{d+1}$ as $Y(d+1) = [y_{d+1}^{(1)}, \ldots, y_{d+1}^{(N)}]$, where $c_{d+1}^{(m)}$ is either a background nucleotide or the start position of a motif site in $S^{(m)}$ for $m = 1, \ldots, N$. We sample the hidden state $Y(d) = [y_d^{(1)}, \ldots, y_d^{(N)}]$ of $a_d$ with probability proportional to $f_d(Y(d)) Tr(Y(d+1) | Y(d), r, t)$ subject to constraints (i) and (ii) of the c-HMM. If the emission components of $a_d$ are sampled as background, we move to $(d-1)$ and repeat to sample $Y(d-1)$. If the emission components of $a_d$ are sampled as state $M$, we further sample the motif type (i.e., the decomposed states $M_0, M_1, \ldots, M_K$) with probabilities proportional to the $K+1$ terms of $f_d(Y(d))$ in (9) for $k \in \{0, 1, \ldots, K\}$. Given the imputed value of $Y(d)$, we set $Y(d - w_k + i) = Y(d)$ for $i = 1, \ldots, w_k - 1$ following constraints (iii) and (ii). Then we move to the point $(d - w_k)$ and repeat the sampling procedure for $Y(d-w_k)$. If $|EC(d)| \geq 2$, we also sample associated ancestral nucleotides and evolutionary bonds according to the calculations in equation (11) to equation (13). Suppose the current emission components are background nucleotides ($B$ or $M_0$). We sample the ancestral nucleotide $Z_d$ with probability

$$P(Z_d = z) \propto \theta_0^{(0)}(z) \prod_{m \in EC(d)} \Phi_b(z, x_d^{(m)}),$$

for $z \in \{A, C, G, T\}$. If the current emission components are binding sites of motif $k$, we sample each base of the ancestral binding site $Z_{d-w_k+1} \cdots Z_d$ independently with probabilities

$$P(Z_{d-w_k+i} = z) \propto \Theta_{ki}(z) \prod_{m \in EC(d)} P(x_{d-w_k+i}^{(m)} | z, \mu_f, \Theta_{ki}),$$

for $z \in \{A, C, G, T\}$ and $i = 1, 2, \ldots, w_k$, where $P(x_{d-w_k+i}^{(m)} | z, \mu_f, \Theta_{ki})$ is given by (13). Given the ancestral binding site, we update the evolutionary bond between the ancestral and current binding sites for each base independently. Denote the evolutionary bond between $x_{d-w_k+i}^{(m)}$ [$m \in EC(d)$] and $Z_{d-w_k+i}$ by $v^{(m)}$. We connect the bond with probability

$$P(v^{(m)} = 1) = \frac{(1 - \mu_f) \cdot \mathbf{1}(x_{d-w_k+i}^{(m)} = Z_{d-w_k+i})}{(1 - \mu_f) + \mu_f \Theta_{ki}(Z_{d-w_k+i})};$$

otherwise the bond will be broken.

**A.4. Sample $\Psi$ given $[Y, Z, V]$ and $A$.** Let us return to the graphical representation of the c-HMM as illustrated in Figure 1(B). In this conditional sampling step, the hidden states, the ancestral nucleotides of coupled nodes and the evolutionary bonds associated with aligned TFBSs are given. We sample $r$ from $Beta(C_{BM} + 1, C_{BB} + 1)$, where $C_{BM}$ and $C_{BB}$ are the numbers of transitions from $B$ to $M$ and from $B$ to $B$, respectively. Denote the numbers of states $B$, $M$ and $M_k$ by $|B|$, $|M|$ and $|M_k|$ for $k = 0, 1, \ldots, K$, respectively ($|M| = \sum_{k=0}^{K} |M_k|$). The conditional posterior distribution of $q = [q_0, q_1, \ldots, q_K]$ is $Dir(|M_0| + 1, \ldots, |M_K| + 1)$. We update the ancestral background distribution $\theta_0^{(0)}$ by $Dir(C_B^{(0)} + \mathbf{1})$, where $C_B^{(0)}$ is the count vector of the imputed ancestral background nucleotides and $\mathbf{1}$ is a vector of 1's of length 4. Denote by $C_i, C_s$ and $C_v$ the numbers of identities, transitions and transversions from the ancestral to the current aligned background nucleotides, respectively. Then we sample $\phi_b = [1 - \mu_b, \alpha, 2\beta]$ from $Dir(C_i + 1, C_s + 1, C_v + 1)$. For all the aligned TFBSs with their imputed ancestors, we count the numbers of broken and connected evolutionary bonds, $|V_0|$ and $|V_1|$, respectively, and sample $\mu_f$ from $Beta(|V_0| + 1, |V_1| + 1)$. The sufficient statistic for $\Theta_{ki}$ ($k = 1, \ldots, K; i = 1, \ldots, w_k$) has three components, (1) the count vector of unaligned current sites, $C_{ki}^g$, (2) the count vector of ancestral sites, $C_{ki}^z$, and (3) the count vector of aligned descendant sites with a broken evolutionary bond, $C_{ki}^b$, since each of them is an independent sample from $\Theta_{ki}$ under our model. Then the conditional posterior distribution of $\Theta_{ki}$ is $Dir(C_{ki} + \mathbf{1})$, where $C_{ki} = C_{ki}^g + C_{ki}^z + C_{ki}^b$.

A Metropolis–Hastings step is implemented to update motif widths. We illustrate our method by one motif as an example. Given the current width $w$ and all sampled sites of this motif, we propose to increase or decrease one base at their left or right ends with equal probability. After choosing one of the four possibilities, the problem is equivalent to a model selection problem: The nucleotides observed in the selected positions are generated from the background ($H_0$) or from a motif column ($H_1$). If $H_1$ is true, denote the weight vector of the motif column by $\Theta_w$ and its sufficient statistic by $C_w$ calculated as described in the previous paragraph for any $\Theta_{ki}$. Before calculating $C_w$, one needs to sample the associated evolutionary bonds $V^w$, with $|V_0^w|$ and $|V_1^w|$ denoting the numbers of broken and connected bonds. Under $H_0$, we denote by $C_b = [c_b^{(0)}, c_b^{(1)}, \ldots, c_b^{(N)}]$ the count vectors of ancestral nucleotides ($c_b^{(0)}$) and current unaligned nucleotides of different orthologs ($c_b^{(m)}; m = 1, \ldots, N$) in the selected positions. Denote by $C_t$ the substitution count matrix from an ancestral base to its descendant bases. Then we calculate the posterior odds of $H_1$ with $V^w$ over $H_0$ by

$$\begin{aligned}
&\frac{P(H_1, V^w | \mu_f, Y, Z, A, S)}{P(H_0 | \theta_0, \Phi_b, Y, Z, A, S)} \\
&= \frac{\pi(H_1)}{\pi(H_0)} \frac{\int P(\Theta_w, V^w, C_w | \mu_f, H_1)\, d\Theta_w}{P(C_b, C_t | \theta_0, \Phi_b, H_0)} \\
&= \frac{\pi(H_1)}{\pi(H_0)} \frac{\mu_f^{|V_0^w|}(1-\mu_f)^{|V_1^w|} \int (\Theta_w)^{C_w} \Gamma(4)\, d\Theta_w}{(\theta_0)^{C_b}(\Phi_b)^{C_t}} \\
&= \frac{\pi(H_1)}{\pi(H_0)} \frac{\mu_f^{|V_0^w|}(1-\mu_f)^{|V_1^w|} \Gamma(C_w + \mathbf{1})\Gamma(4)}{\prod_{m=0}^N (\theta_0^{(m)})^{c_b^{(m)}} (\Phi_b)^{C_t} \Gamma(|C_w| + 4)},
\end{aligned}$$
(15)

where we define $R^C = \prod_{i,j} R_{ij}^{C_{ij}}$ and $\Gamma(R) = \prod_{i,j} \Gamma(R_{ij})$ with $R$ and $C$ being matrices (vectors) of same size, $|C_w|$ is the total counts in $C_w$, and the ratio of $\pi(H_1)$ over $\pi(H_0)$ is determined by the Poisson prior for the motif width. In each iteration, we always propose to flip the two hypotheses. The proposal from $H_0$ to $H_1$ involves proposing evolutionary bonds for all the aligned nucleotides that are identical to their ancestors. We propose to break each of these bonds independently with probability $\mu_f$. Under these proposals, the Metropolis–Hastings ratio is

$$
\begin{aligned}
R_{\mathrm{MH}} &= \frac{P(H_1, V^w | \mu_f, Y, Z, A, S)}{P(H_0 | \theta_0, \Phi_b, Y, Z, A, S)} \frac{Q(H_0 | H_1, V^w)}{Q(H_1, V^w | H_0)} \\
&= \frac{P(H_1, V^w | \mu_f, Y, Z, A, S)}{P(H_0 | \theta_0, \Phi_b, Y, Z, A, S)} \frac{1}{\mu_f^{|V_0^w| - n_s} (1 - \mu_f)^{|V_1^w|}} \\
&= \frac{\pi(H_1)}{\pi(H_0)} \frac{\mu_f^{n_s} \Gamma(C_w + \mathbf{1}) \Gamma(4)}{\prod_{m=0}^{N} (\theta_0^{(m)})^{c_b^{(m)}} (\Phi_b)^{C_t} \Gamma(|C_w| + 4)},
\end{aligned}
\tag{16}
$$

where $n_s$ is the number of aligned nucleotides that are different from their ancestors, and $Q$ stands for the proposal probabilities. From the definition of an evolutionary bond, we always have $n_s \leq |V_0^w|$.

**A.5. Sample $A$ given $\Psi$.** To consider the uncertainty in multiple alignments, we update $A$ by its marginal posterior distribution given the current parameters, that is, we want to sample from $P(A|\Psi, S)$. A Metropolis–Hastings step is implemented for this conditional sampling step. Suppose the current alignment is $A$. We propose a new alignment $A^*$ from an ordinary multiple sequence alignment procedure based on an HMM [Baldi et al. (1994), Krogh et al. (1994)], denoted by MA-HMM so as to distinguish from c-HMM. In MA-HMM, each sequence is aligned to a profile (or a hidden sequence template) based on an HMM with three states, aligned, insertion and deletion. In our proposal, the transition matrix between the three states is fixed by prior expectations as

$$
D = \begin{bmatrix} 0.998 & 0.001 & 0.001 \\ 0 & 0.998 & 0.002 \\ 0.025 & 0.025 & 0.95 \end{bmatrix},
$$

where states are ordered as deletion, insertion and aligned. The rationale for selecting these values is that we expect to start an aligned block every 500 bps ($D_{11} = D_{22} = 0.998$) and that the average length of an aligned block is 20 bps ($D_{33} = 0.95$). These values serve as the default parameters for all the results presented in this article. We use the current neutral substitution matrix as the emission probabilities from the profile (ancestral sequence $Z$) to a current aligned nucleotide. Ancestral and unaligned nucleotides are emitted from their respective background models $\theta_0^{(0)}, \theta_0^{(1)}, \ldots, \theta_0^{(N)}$. Denote the probability of proposing the alignment $A^*$ by $Q(A^* | \theta_0, \Phi_b, Z, S)$ under the MA-HMM. Then the Metropolis–Hastings ratio is

$$
\frac{P(A^* | \Psi, S)}{P(A | \Psi, S)} \frac{Q(A | \theta_0, \Phi_b, Z, S)}{Q(A^* | \theta_0, \Phi_b, Z, S)} = \frac{P(S | A^*, \Psi)}{P(S | A, \Psi)} \frac{Q(S, Z | A, \theta_0, \Phi_b)}{Q(S, Z | A^*, \theta_0, \Phi_b)},
\tag{17}
$$

where $P(S|A, \Psi)$ and $P(S|A^*, \Psi)$ are calculated by (14) through the recursive forward summations. $Q(S, Z | A, \theta_0, \Phi_b)$ and $Q(S, Z | A^*, \theta_0, \Phi_b)$ are the

probabilities of observing $S$ and $Z$ given alignments $A$ and $A^*$ under the MA-HMM, respectively, that is,

$$(18) \quad Q(S,Z|A,\theta_0,\Phi_b) = \prod_{m=0}^{N} (\theta_0^{(m)})^{c_b^{(m)}} (\Phi_b)^{C_t},$$

where $c_b^{(m)}$ and $C_t$ are defined similarly to those in (15) but for all the sequence positions. Note that the prior probabilities of an alignment are identical in both c-HMM and MA-HMM, and thus are canceled at the R.H.S. of equation (17).

**A.6. A collapsed sampler.** Suppose the data set contains $n$ genes. Denote the missing data, including $Y, Z, A$ and $V$, for the $i$th gene by $D_{\text{mis},i}$, and let $S_i = \{S_i^{(m)}\}_{m=1}^N$ be the orthologous sequences of the $i$th gene, $i = 1, 2, \ldots, n$. Since collapsing random components in the Gibbs sampler usually results in more efficient sampling schemes [Liu, Wong and Kong (1994), Liu (1994)], we implement a collapsed MultiModule to sample from $P(D_{\text{mis},1}, \ldots, D_{\text{mis},n}|S)$ by iteratively scanning each gene. For the $i$th gene, given the current imputed values of

$$D_{\text{mis},[-i]} = \{D_{\text{mis},1}, \ldots, D_{\text{mis},i-1}, D_{\text{mis},i+1}, \ldots, D_{\text{mis},n}\},$$

we first estimate the parameters by their conditional posterior means,

$$(19) \quad \hat{\Psi}_{[-i]} = E(\Psi|D_{\text{mis},[-i]}, S_{[-i]}),$$

where $S_{[-i]} = \{S_1, \ldots, S_{i-1}, S_{i+1}, \ldots, S_n\}$. Then we sample the missing data for the $i$th gene from

$$(20) \quad D^*_{\text{mis},i} \sim P(D_{\text{mis},i}|\hat{\Psi}_{[-i]}, S_i).$$

Equation (19) can be easily calculated from the conditional posterior distributions of different parameters (Section A.4), and equation (20) is exactly the same sampling procedure as described in Sections A.3 and A.5 taking $\hat{\Psi}_{[-i]}$ as the current parameters. We want to emphasize that, although this is only an approximate way to collapse all the parameters, it indeed improves the convergence of the Gibbs sampler. The simulation studies were performed with this collapsed version of MultiModule.

**Acknowledgments.** The authors thank Professors Arthur P. Dempster and Jun S. Liu for their helpful comments.

## REFERENCES

Baldi, P., Chauvin, Y., Hunkapiller, T. and McClure, M. A. (1994). Hidden Markov models of biological primary sequence information. *Proc. Natl. Acad. Sci. USA* **91** 1059–1063.

Berman, B. P., Nibu, Y., Pfeiffer, B. D., Tomancak, P., Celniker, S. E., Levine, M., Rubin, G. M. and Eisen, M. B. (2002). Exploiting transcription factor binding site clustering to identify *cis*-regulatory modules involved in pattern formation in the Drosophila genome. *Proc. Natl. Acad. Sci. USA* **99** 757–762.

Boyer, L. A., Lee, T. I., Cole, M. F., Johnstone, S. E., Levine, S. S., Zucker, J. P. et al. (2005). Core transcriptional regulatory circuitry in human embryonic stem cells. *Cell* **122** 947–956.

Brudno, M., Do, C. B., Cooper, G. M., Kim, M. F., Davydov, E., NISC Comparative Sequencing Program, Green, E. D., Sidow, A. and Batzoglou, S. (2003). LAGAN and Multi-LAGAN: Efficient tools for large-scale multiple alignment of genomic DNA. *Genome Res.* **13** 721–731.

Bussemaker, H. J., Li, H. and Siggia, E. D. (2000) Building a dictionary for genomes: Identification of presumptive regulatory sites by statistical analysis. *Proc. Natl. Acad. Sci. USA* **97** 10096–10100. MR1773836

Felsenstein, J. (1981). Evolutionary trees from DNA sequences: A maximum likelihood approach. *J. Mol. Evol.* **17** 368–376.

Frith, M. C., Hansen, U. and Weng, Z. (2001). Detection of *cis*-element clusters in higher eukaryotic DNA. *Bioinformatics* **17** 878–889.

Frith, M. C., Spouge, J. L., Hansen, U. and Weng, Z. (2002). Statistical significance of clusters of motifs represented by position specific scoring matrices in nucleotide sequences. *Nucleic Acids Res.* **30** 3214–3224.

Geyer, C. J. (1991). Markov chain Monte Carlo maximum likelihood. In *Computing Science and Statistics*: *Proceedings of the 23rd Symposium on the Interface* (E. M. Keramigas, ed.) 156–163. Interface Foundation, Fairfax, VA.

Gupta, M. and Liu, J. S. (2005). *De novo cis*-regulatory module elicitation for eukaryotic genomes. *Proc. Natl. Acad. Sci. USA* **102** 7079–7084.

Hampson, S., Kibler, D. and Baldi, P. (2002). Distribution patterns of over-represented kmers in non-coding yeast DNA. *Bioinformatics* **18** 513–528.

Jensen, S. T., Liu, X. S., Zhou, Q. and Liu, J. S. (2004). Computational discovery of gene regulation binding motifs: A Bayesian perspective. *Statist. Sci.* **19** 188–204. MR2082154

Johnson, D. S., Zhou, Q., Yagi, K., Satoh, N., Wong, W. H. and Sidow, A. (2005). *De novo* discovery of a tissue-specific gene regulatory module in a Chordate. *Genome Res.* **15** 1315–1324.

Kou, S. C., Zhou, Q. and Wong, W. H. (2006). Equi-energy sampler with applications in statistical inference and statistical mechanics (with discussion). *Ann. Statist.* **34** 1581–1652. MR2283711

Krogh, A., Brown, M., Mian, L. S., Sjöander, K. and Haussler, D. (1994). Hidden Markov models in computational biology: Applications to protein modeling. *J. Mol. Biol.* **235** 1501–1531.

Lawrence, C. E., Altschul, S. F., Boguski, M. S., Liu, J. S., Neuwald, A. F. and Wooton, J. C. (1993). Detecting subtle sequence signals: A Gibbs sampling strategy for multiple alignment. *Science* **262** 208–214.

Lawrence, C. E. and Reilly, A. A. (1990). An expectation maximization (EM) algorithm for the identification and characterization of common sites in unaligned biopolymer sequences. *Proteins* **7** 41–51.

Li, X. and Wong, W. H. (2005). Sampling motifs on phylogenetic trees. *Proc. Natl. Acad. Sci. USA* **102** 9481–9486. MR2168717

Liu, J. S. (1994). The collapsed Gibbs sampler in Bayesian computations with applications to a gene regulation problem. *J. Amer. Statist. Assoc.* **89** 958–966. MR1294740

Liu, J. S., Neuwald, A. F. and Lawrence, C. E. (1995). Bayesian models for multiple local sequence alignment and Gibbs sampling strategies. *J. Amer. Statist. Assoc.* **90** 1156–1170.

Liu, J. S., Wong, W. H. and Kong, A. (1994). Covariance structure of the Gibbs sampler with applications to the comparisons of estimators and augmentation schemes. *Biometrika* **81** 27–40. MR1279653

Liu, X. S., Brutlag, D. L. and Liu, J. S. (2002). An algorithm for finding protein-DNA binding sites with applications to chromatin immunoprecipitation microarray experiments. *Nat. Biotech.* **20** 835–839.

Liu, Y., Liu, X. S., Wei, L., Altman, R. B. and Batzoglou, S. (2004). Eukaryotic regulatory element conservation analysis and identification using comparative genomics. *Genome Res.* **14** 451–458.

Loots, G. G., Locksley, R. M., Blankespoor, C. M., Wang, Z. E., Miller, W., Rubin, E. M. and Frazer, K. A. (2000). Identification of a coordinate regulator of interleukins 4, 13, and 5 by cross-species sequence comparisons. *Science* **288** 136–140.

Moses, A. M., Chiang, D. Y. and Eisen, M. B. (2004). Phylogenetic motif detection by expectation–maximization on evolutionary mixtures. *Pac. Smp. Biocomput.* **9** 324–335.

Prakash, A., Blanchette, M., Sinha, S. and Tompa, M. (2004). Motif discovery in heterogeneous sequence data. *Pac. Smp. Biocomput.* **9** 348–359.

Roth, F. R., Hughes, J. D., Estep, P. E. and Church, G. M. (1998). Finding DNA regulatory motifs within unaligned noncoding sequences clustered by whole genome mRNA quantization. *Nat. Biotech.* **16** 939–945.

Sanchez, L. and Thieffry, D. (2001). A logical analysis of the Drosophila gap-gene system. *J. Theor. Biol.* **211** 115–141.

Schneider, T. D. and Stephens, R. M. (1990). Sequence logos: A new way to display consensus sequences. *Nucleic Acids Res.* **18** 6097–6100.

Siddharthan, R., Siggia, E. D. and van Nimwegen, E. (2005). PhyloGibbs: A Gibbs sampling motif finder that incorporates phylogeny. *PLoS Comput. Biol.* **1** e67.

Sinha, S., Blanchette, M. and Tompa, M. (2004). PhyME: A probabilistic algorithm for finding motifs in sets of orthologous sequences. *BMC Bioinformatics* **5** 170.

Sinha, S. and Tompa, M. (2002). Discovery of novel transcription factor binding sites by statistical overrepresentation. *Nucleic Acids Res.* **30** 5549–5560.

Sinha, S., van Nimwegen, E. and Siggia, E. D. (2003). A probabilistic method to detect regulatory modules. *Bioinformatics* **19** (Suppl.) i292–i301.

Stormo, G. D. and Hartzell, G. W. (1989). Identifying protein-binding sites from unaligned DNA fragments. *Proc. Natl. Acad. Sci. USA* **86** 1183–1187.

Thompson, W., Palumbo, M. J., Wasserman, W. W., Liu, J. S. and Lawrence, C. E. (2004). Decoding human regulatory circuits. *Genome Res.* **14** 1967–1974.

Wang, T. and Stormo, G. D. (2003). Combining phylogenetic data with co- regulated genes to identify regulatory motifs. *Bioinformatics* **19** 2369–2380.

Wasserman, W. W., Palumbo, M., Thompson, W., Fickett, J. W. and Lawrence, C. E. (2000). Human–mouse genome comparisons to locate regulatory sites. *Nat. Genet.* **26** 225–228.

Wingender, E., Chen, X., Hehl, R., Karas, H., Liebich, I., Matys, V., Meinhardt, T., Pruss, M., Reuter, I. and Schacherer, F. (2000). TRANSFAC: An integrated system for gene expression regulation. *Nucleic Acids Res.* **28** 316–319.

Xie, X., Lu, J., Kulbokas, E. J., Golub, T. R., Mootha, V., Lindblad-Toh, K., Lander, E. S. and Kellis, M. (2005). Systematic discovery of regulatory motifs in human promoters and 3' UTRs by comparison of several mammals. *Nature* **434** 338–345.

Yuh, C. H., Bolouri, H. and Davidson, E. H. (1998). Genomic *cis*-regulatory logic: Experimental and computational analysis of a sea urchin gene. *Science* **279** 1896–1902.

Zhou, Q. and Wong, W. H. (2004). CisModule: *De novo* discovery of cis-regulatory modules by hierarchical mixture modeling. *Proc. Natl. Acad. Sci. USA* **101** 12114–12119.

Department of Statistics  
UCLA  
8125 Math Sciences Bldg.  
Box 951554  
Los Angeles, California 90095  
USA  
E-mail: zhou@stat.ucla.edu

Departments of Statistics,  
 Health Research and Policy  
 and Biological Sciences  
Stanford University  
390 Serra Mall  
Stanford, California 94305  
USA  
E-mail: whwong@stanford.edu